\documentclass[11pt]{article}

\usepackage{sao1}
\usepackage{graphicx}
\usepackage{amsmath,graphicx,amssymb}

\begin{document}

\title{Magnetic Chemically Peculiar Stars with Unsteady Periods }

\author{Mikul\'a\v sek~Z. \inst{1,}\inst{2}
       \and
       Krti\v{c}ka~J. \inst{1}
       \and
       Jan\'ik~J. \inst{1}
       \and
       Zverko~J. \inst{3}
       \and
       \v Zi\v z\v novsk\'y~J. \inst{3}
       \and
       Zv\v{e}\v{r}ina~P. \inst{1}
       \and
       Zejda~M. \inst{1}
        }
\institute{
           Department of Theoretical Physics and Astrophysics,
           Masaryk University, Brno, Czech Republic
       \and
       Observatory and Planetarium of J. Palisa, V\v SB~--- Technical
           University, Ostrava, Czech Republic
       \and
           Astronomical Institute, Slovak Academy of Sciences, Tatransk\'{a}
           Lomnica, Slovak Republic
        }
\maketitle

\begin{abstract}
Photometrically and spectroscopically variable chemically peculiar
(CP) stars are the optimum laboratories for testing the rotational
evolution of main sequence stars. A vast majority of well--studied
CP stars have quite steady rotational periods (e.\,g.\ SrCrEu star
CQ\,UMa). However, there are several CP stars that exhibit
apparent period variations. The origin of the period variations is
unclear in many cases. We describe the observed period variations
of several individual CP stars, especially V901\,Ori,
$\sigma$\,Ori\,E, HR\,7355, CU\,Vir, SX\,Ari, and EE\,Dra. The CP
stars with unsteady periods now represent a very diverse group
with dissimilar O--C diagrams and time scales. We also discuss the
causes of the period changes found and a possible cyclicity or
chaoticity of them.
\end{abstract}

\section{Introduction}

Stars originate from a gravitational collapse of dense parts of
molecular clouds. Every star, besides its matter, inherits a
fraction of the angular momentum of the mother cloud, consequently
each star does rotate.

Stars spend the prevailing part of their active lifetime as main
sequence objects. During the whole MS stage the stellar angular
momentum is largely conserved. For example, the stellar wind takes
away a significant fraction of the total angular momentum only in
the case of very massive stars. Besides the angular momentum the
rotational period of a main sequence star is also determined by
its instant radius, and the inner distribution of its mass and
angular momentum. The development of the rotational period will be
then gradual on the scale of $10^7$--$10^9$~years.

The evolutionary models corresponding to CP stars show that the
equatorial rotational velocity remains practically constant during
the MS epoch (see e.\,g.\ Meynet~\& Maeder, 2000). How can we test
it?

Global data on the rotational period of a MS star and its
evolution can be derived from the $v \sin i$ rotational
broadening. However,the method cannot be applied to an individual
single star, since we do not know the values of its radius and
inclination.

To find out the rotational period changes (if any), a much finer
instrument is needed. Spotty magnetic chemically peculiar (mCP)
stars with a global magnetic field and stable surface structures,
whose periods of light, spectral and magnetic filed variations is
equal to the rotational one, can serve as the best such
instrument.

Combining both the present and archive photometric, spectroscopic
and spectropolarimetric observations collected during many
decades, one can reconstruct the development of the rotation at
least of the outer parts of a star with high accuracy.

Careful period analyses of several dozens of mCP stars have been
done. They confirmed the expectations that the rotational periods
of most of upper MS stars are quite steady. However, a few stars
show period changes, the origin of which has not been completely
understood yet.

\section{Individual Stars}

\subsection{SrCrEu mCP Star CQ~Ursae Majoris}

\begin{figure}
\begin{minipage}[h]{0.5\linewidth}
\center{\includegraphics[width=8.cm,angle=0]{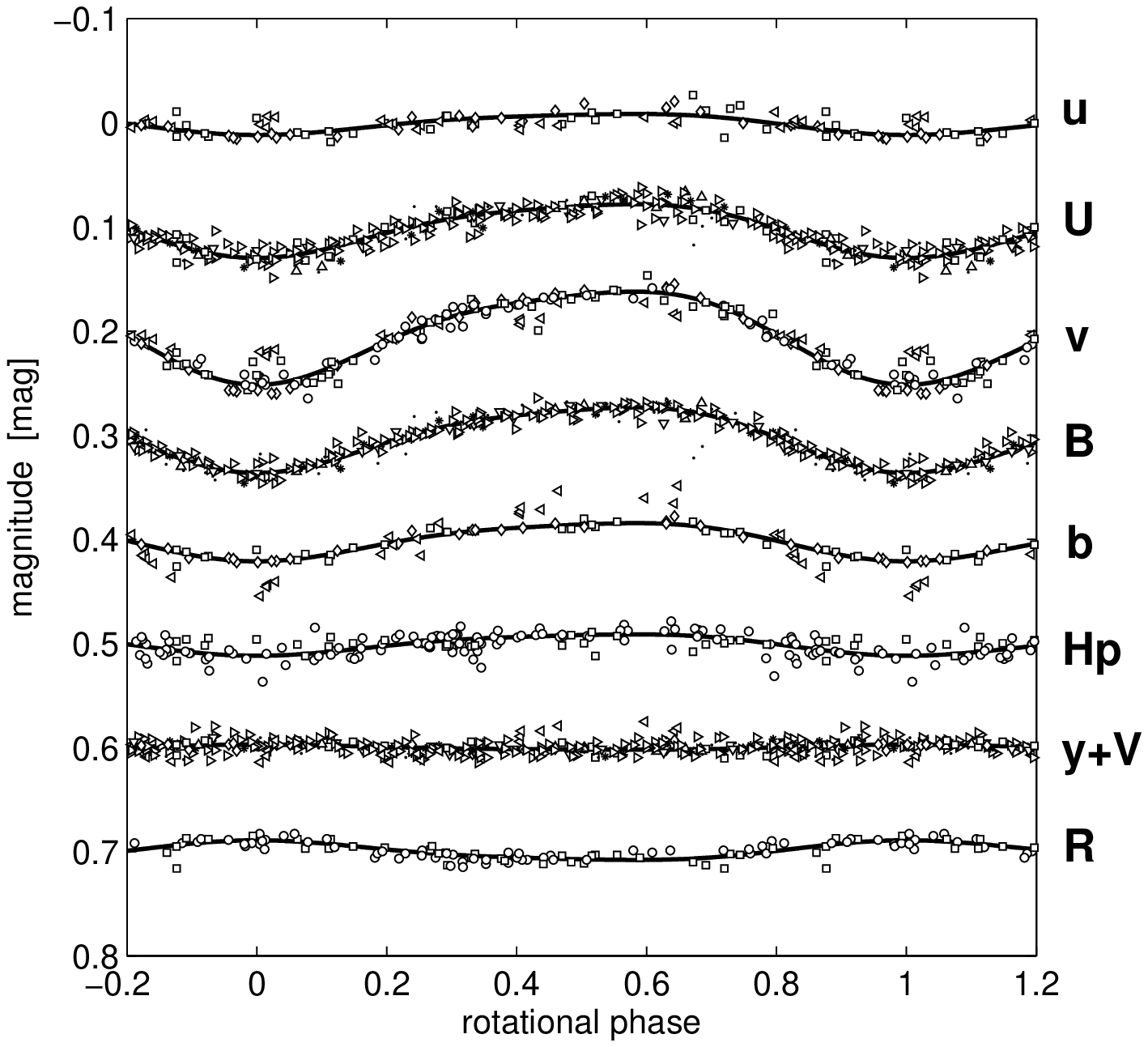}\\ (a)}
\end{minipage}
\begin{minipage}[h]{0.5\linewidth}
\center{\includegraphics[width=8.cm,angle=0]{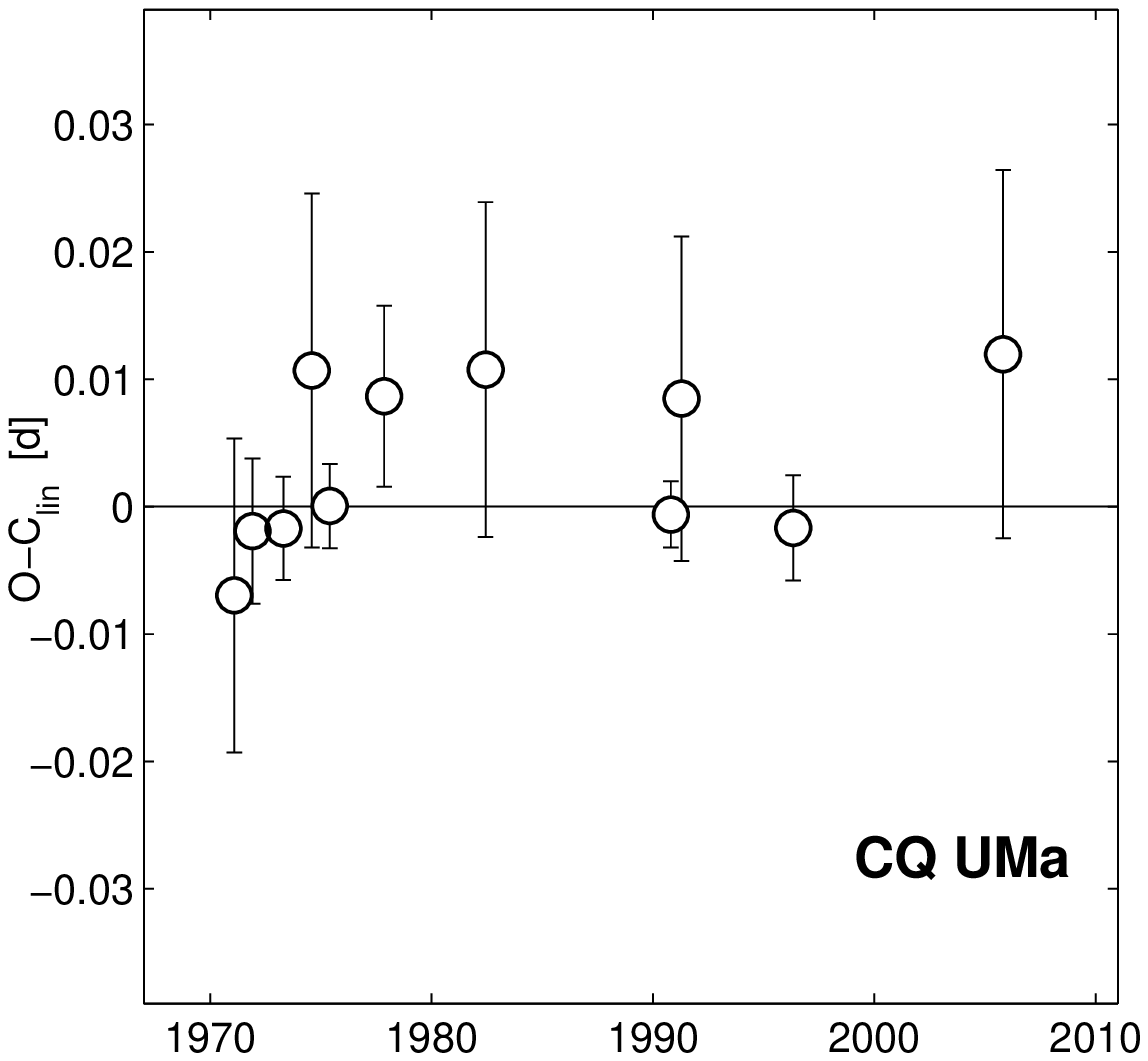}\\ (b)}
\end{minipage}
\caption{(a)~CQ\,UMa light curves in $\mathit{u,\ U,\ v, B,\ b,\
Hp,\ V\!+\!y}$ and $R$--bands. Note the disappearance of
variations in $V\!+\!y$ and the antiphase variations in the
$R$--band. The linear ephemeris: $M_0\!=\!2444384.432$,
$P\!=\!2\fd4499120(27)$. (b)~The time development of the
difference between the observed (O) and calculated (C) times of
the zero phase according to this linear ephemeris depicts the
so--called `O--C diagram'. No trend in the diagram indicates that
the rotational period of the star is constant over the
decades.}\label{CQkrivky}
\end{figure}

As an example of the strictly periodic star we mention is
CQ\,UMa~= HR\,5153~= HD\,119213. This ``cool'' SrCrEu mCP star
displays a prominent variation in the Str\"omgren $v$--band with
antiphase changes in the red band (see Fig.\,\ref{CQkrivky}a).

We used 1365 individual observations collected in 11 various
sources of photometric data that cover a time interval of 42 years
(6262 revolutions of the star). The mean period:
$P\!=\!2\fd4499120(27)$ can be then derived with  the accuracy of
0.23\,s.  The linear fit in  the O--C diagram is depicted in
Fig.\,\ref{CQkrivky}b (the phases of minima are computed for the
$v$--band). The time derivative of the period {\bf is}
$\dot{P}\!=\!(3\pm7)$\,s/cen
 (seconds per century)  that means the period is stable as for
 most of other CP stars. However, there are CP stars showing
 indubitable changes of their periods.

\subsection{He--Strong mCP Star V901\,Orionis}

V901\,Ori~= HD\,37776 is a very young hot star (B2\,IV) residing
in the emission nebula IC\,432, with a complex (quadruple) global
magnetic field (Thompson~\& Landstreet, 1985; Kochukhov et al.,
2011). It can be ranked among the He--strong mCP stars, however,
the light variations are due to the spots of overabundant Si and
He (Krti\v{c}ka et al., 2007).

\begin{figure}[ht!]
\begin{minipage}[h]{0.5\linewidth}
\center{\includegraphics[width=8.5cm,angle=0]{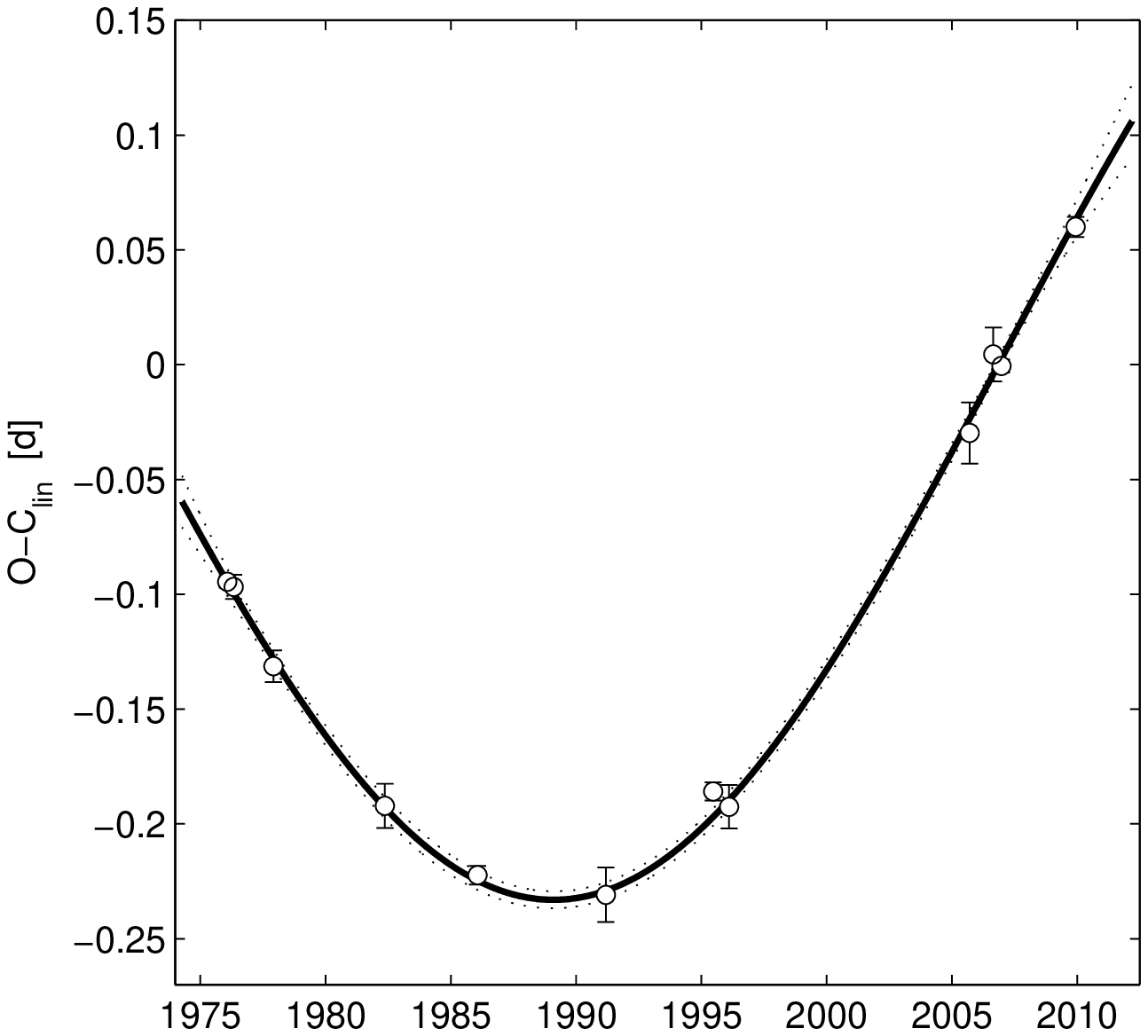}\\ (a)}
\end{minipage}
\begin{minipage}[h]{0.5\linewidth}
\center{\includegraphics[width=8.4cm,angle=0]{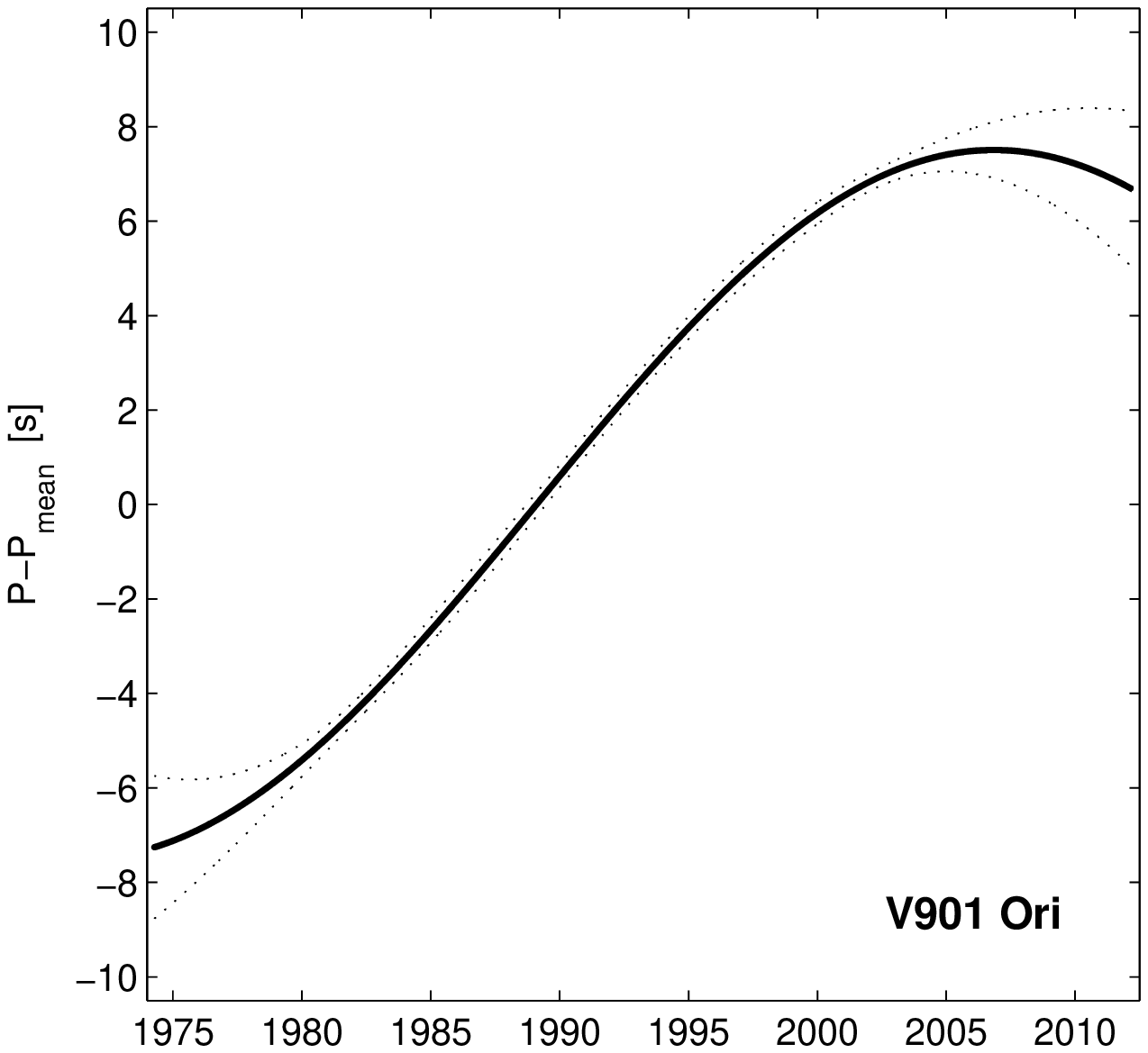}\\ (b)}
\end{minipage}
\caption{(a)~The nonlinear trend in the O--C diagram of V901\,Ori
documents an extraordinarily strong increase of the period in the
time interval of 1976--2005. Now (the end of 2010) the period is
nearly constant. The O--C values were calculated relatively to the
linear ephemeris: $M_0\!=\!2445724.669$, $P\!=\!1\fd5386754$
published by Adelman (1997b). (b)~The dependence of the difference
between the observed and the mean periods (in seconds) on time
does not exclude the possibility of cyclic variations of the
period.}\label{OC_901}
\end{figure}

Using photometry and spectroscopy, Mikul\'a\v{s}ek et al.\ (2008)
proved that the observed period of about $1\fd5387$ has been
gradually changing. The O--C diagram (see Fig.\,\ref{OC_901}a) can
be formally fitted with a smooth curve either of a 4--th order
polynomial or a segment of a cosinusoid. The maximum
increase of the period $\dot{P}_{\rm{max}}\!=\!2.10(16)\cdot
10^{-8}\!=\!66(5)$\,s/cen took place around the year 1989. The
mean increase of the period during the
 recent 35 years is only $\overline{\dot{P}}\!=\!1.7\cdot 10^{-8}=53$\,s/cen.
The value of $\dot{P}$ is now (at the end of 2010)
definitely much smaller: $\dot{P}\!=\!-8(7)\cdot 10^{-9}\!=\!-25(22)$\,s/cen.

Ruling out the light--time effect in a binary star, the precession of
rotational axis, and the evolutionary changes as possible causes of the
period change, we interpret it in terms of braking of the star's
rotation (at least of its surface layers) due to the angular momentum loss through
events in the stellar magnetosphere (Mikul\'a\v{s}ek et al., 2008). However, this
mechanism is unable to explain the possible acceleration of the
rotation nowadays.

\subsection{Helium--Strong mCP Star $\bf\sigma$\,Orionis\,E}

The spectrum of a very young star $\sigma$\,Ori\,E~= HD\,37479 is
a hybrid of a classical He--strong mCP star and a B emission--line
star (Walborn, 1974). The light curves in the optical domain,
namely in the $u$~$(U)$--band, are unusual  for the CP stars: the
narrow and deep minima cannot be explained in terms of photometric
spots on the surface only. A contribution of ``eclipses'' by
magnetospheric ``clouds'' (Landstreet~\& Borra, 1978; Townsend et
al., 2005) must be allowed.

Townsend et al.\ (2010) {discovered recently a smooth rotational
braking in a moderate rate $\dot{P}\!=\!7.7$\,s/cen based on their
$U$ observations obtained within 2004--2009, and the $u$
observations obtained in 1977 by Hesser et al.\ (1977). Townsend
et al.\ (2010) explained the observed spin--down of the star by
the magnetic braking through the line--driven stellar wind.

\subsection{Helium--Strong mCP Star HR\,7355}

The O--C diagram of another He--strong, very rapidly rotating mCP star
HR\,7355~= HD\,180182 $(P\!=\!0\fd5214)$ with emission lines is
similar to the above discussed stars. This suggest HR\,7355 might be
also a spin--down of hot mCP star (Mikul\'a\v{s}ek et al., 2010). The only puzzling
aspect is the rather advanced age of the star (20\,Myr).

\begin{figure}
\begin{minipage}[h]{0.5\linewidth}
\center{\includegraphics[width=8.5cm,angle=0]{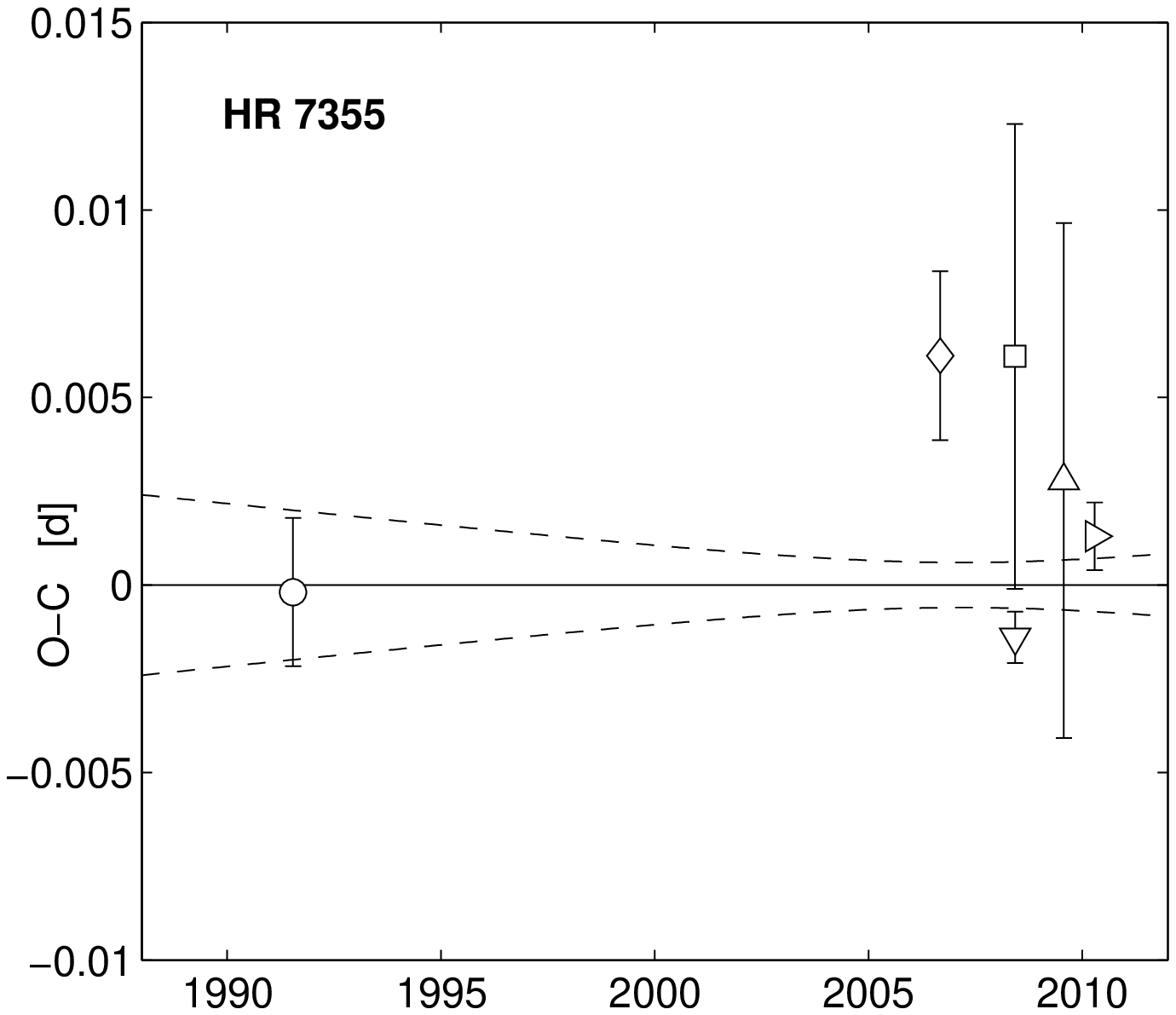}\\ (a)}
\end{minipage}
\begin{minipage}[h]{0.5\linewidth}
\center{\includegraphics[width=8.3cm,angle=0]{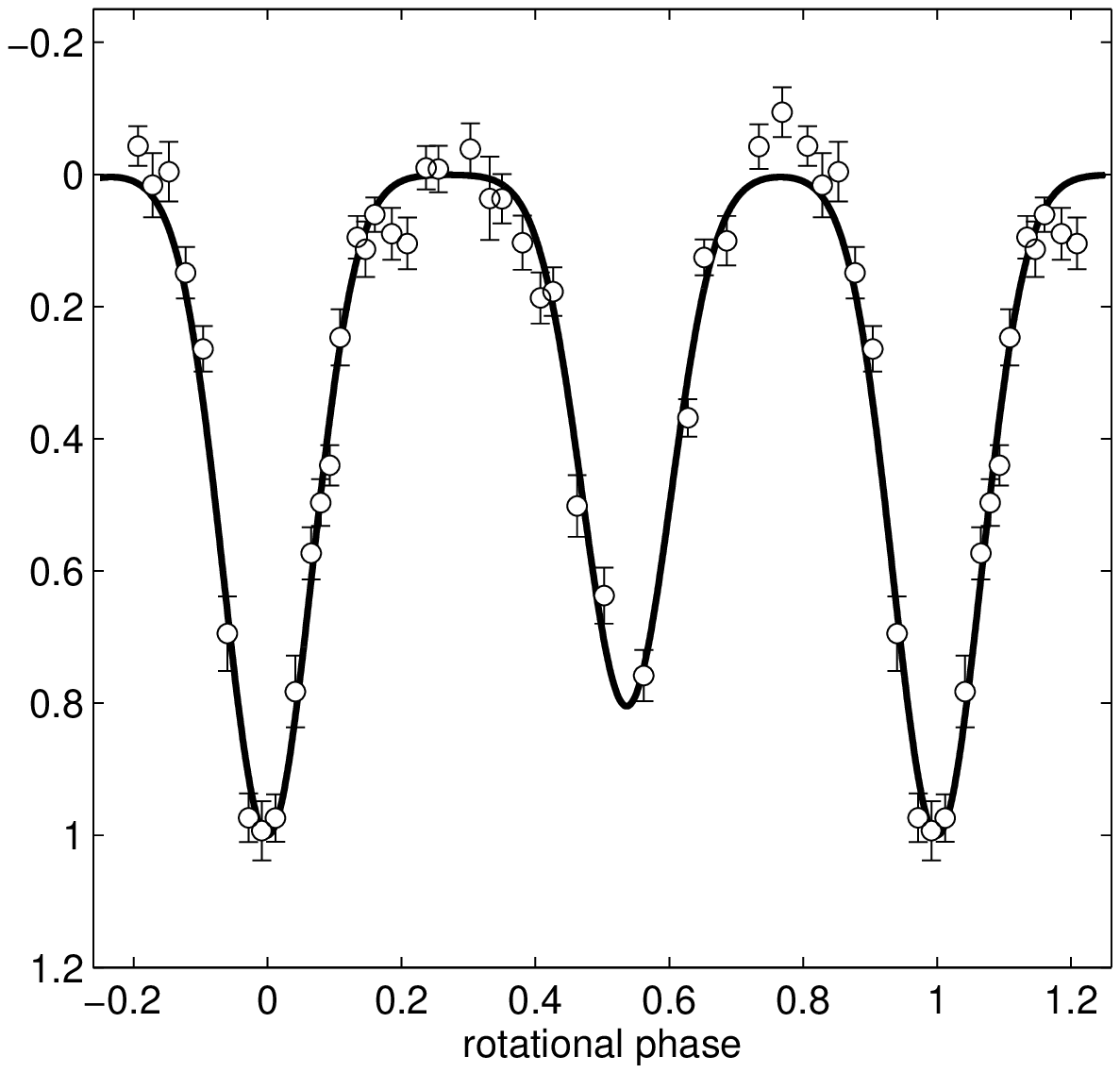}\\ (b)}
\end{minipage}
\caption{(a)~The up--to--date O--C diagram of HR\,7355 does not
indicate any period changes. The open circle: Hipparcos
observations, the diamond: ASAS measurements; the square and
$\triangle$:~observations published by Mikul\'a\v{s}ek et al.\
(2010);
 $\nabla$:~the $R$ measurements of Oksala et al.\ (2010);
$\rhd$:~our unpublished $UBV$ observations. (b)~The light curve of HR\,7355
represented by normal points.  The narrow minima cannot be merely
due to photometric spots as it is common in the CP stars.}
\label{OC_7355}
\end{figure}

However, the recent revision of the ASAS data on HR\,7355
(Pojma\'nski et al., 2010) and the two new extended sets of
photometry, obtained recently by Oksala et al.\ (2010) and our
group, ruled out this suspicion reliably. The latest O--C diagram,
Fig.\,\ref{OC_7355}a, does not indicate any change of the period.
The new light curves show the star is an ``elder sister'' of
$\sigma$\,Ori\,E with eclipses (as it was proposed in Rivinius et
al., 2008), but no braking (now).

\subsection{Silicon mCP Star CU\,Virginis}

The famous very fast--rotating silicon mCP star CU\,Vir~=
HD\,124224~= HR\,5313 may show another type of period changes.
Amplitudes of the light and spectral variations (He\,I, Si\,II,
H\,I, and other) are relatively large.  CU\,Vir is among the most
frequently studied mCP stars, consequently, its behaviour is
reliably documented. Moreover, CU\,Vir is a unique main sequence
radio pulsar (Trigilio et al., 2008; Ravi et al., 2010).

\begin{figure}[ht!]
\begin{minipage}[h]{0.5\linewidth}
\center{\includegraphics[width=8.4cm,angle=0]{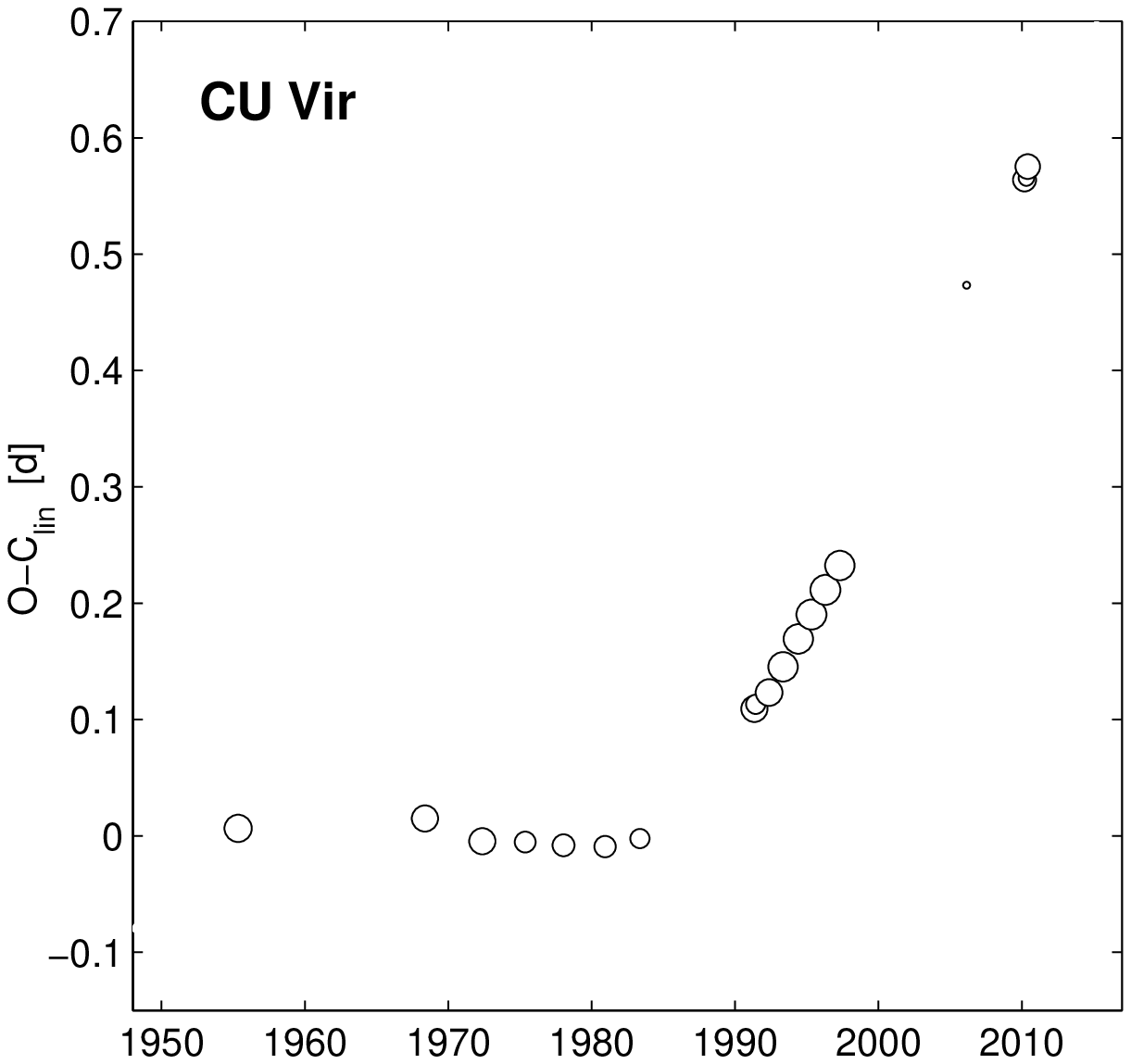}\\ (a)}
\end{minipage}
\begin{minipage}[h]{0.5\linewidth}
\center{\includegraphics[width=8.5cm,angle=0]{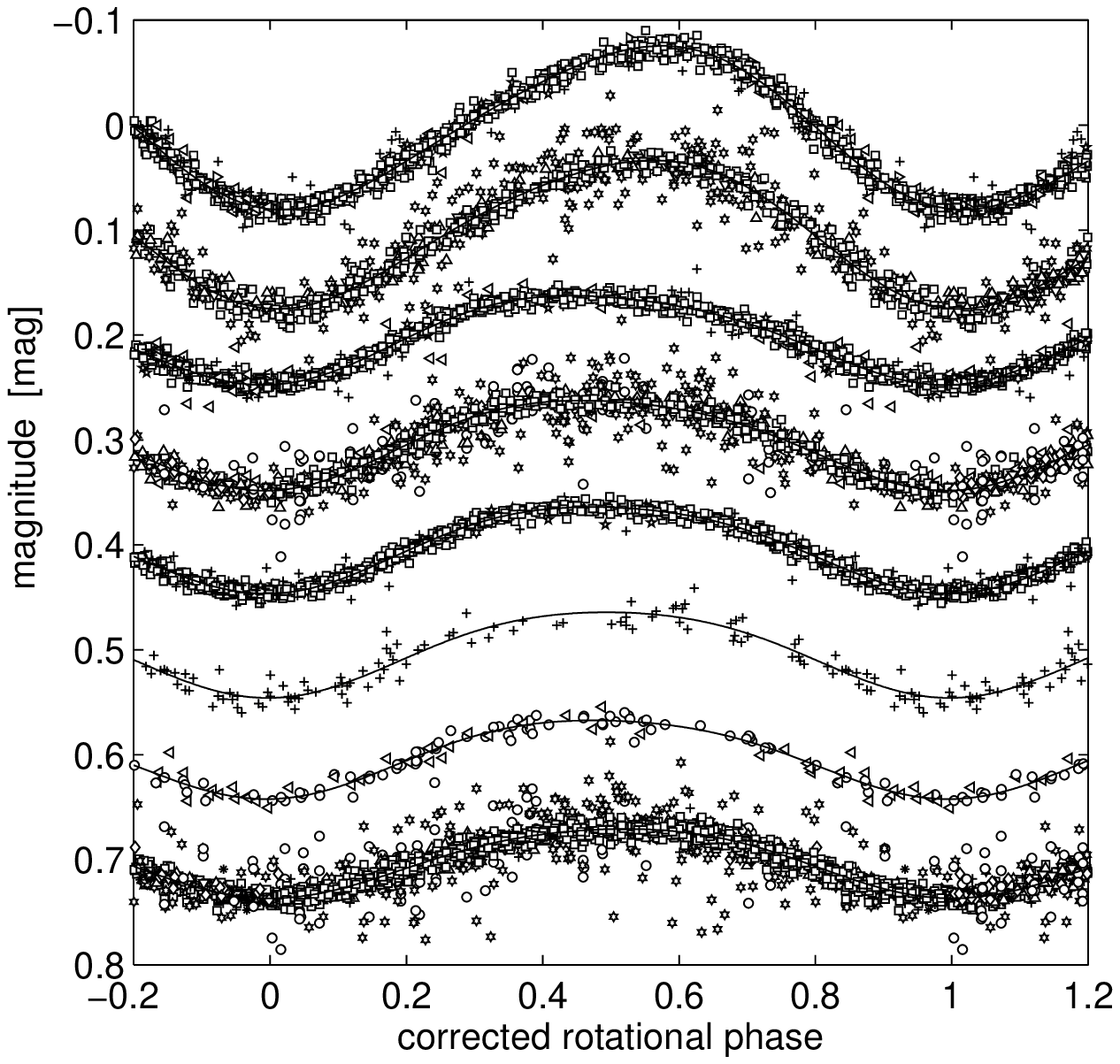}\\ (b)}
\end{minipage}
\caption{(a) The O--C diagram of CU\,Vir for the times of light
minima derived from the available photometries (Hardie, 1958;
Blanco~\& Catalano, 1971; Winzer, 1974; Molnar~\& Wu, 1978;
Pyper~\& Adelman, 1985; Sokolov, 2000; Pyper et al., 1998;
Pojma\'nski et al., 2001) and our new unpublished data according
to the ephemeris in Pyper et al.\ (1998): $M_0\!=\!2435178.6417$,
$P\!=\!0\fd5206778$. The size of an open circle correspond to the
weight of the value, standard accuracy of the value is 0\fd0025.
Two or three linear segments can fit the course. Eventually, a
more complex smooth function (Pyper et al., 1998; Pyper~\&
Adelman, 2004) can be used. (b)~The light curves in $\mathit{u,\
U,\ v,\ B,\ b,\ W\beta,\ Hp}$ and $V\!+\!y$--bands (arranged from
top to bottom) were assembled from 7097 individual photometric
observations. Note the gradual change of the shape of the light
curves with the effective wavelength of a particular colour band.
The points, tightly adjoined the light curves, corrected for the
change of period, show that shapes of the individual light curves
are constant over the past sixty years.}\label{OC_CU}
\end{figure}

Pyper et al.\ (1998) discovered an abrupt increase of the period
from $0\fd5206778$ to $0\fd52070854$ that occurred approximately
in 1984 and Pyper~\& Adelman (2004) discussed two possible
scenarios of the explanation of the observed O--C diagram, namely
a continually changing period or two constant periods.

The mean deceleration of the period of CU\,Vir during the past 60
years is $\overline{\dot{P}}\!=\!2.4\cdot 10^{-9}\!=\!7.6$\,s/cen.
The estimated maximum increase (near 1984) is
$\dot{P}_{\rm{ex}}\!=\!5.7\cdot 10^{-9}\!=\!18$\,s/cen. Our
photometric and spectroscopic observations obtained in 2009--2010
indicate that the period is now constant.

Presently, we are recalculating the whole O--C diagram using all
available data, containing phase data. The results will be
published in forthcoming papers.

The shapes of the light curves of V901\,Ori and CU\,Vir, the
prototypes of mCP stars with large period variations are
non--variable, thus excluding  precession as the cause of observed
period changes (for details see Mikul\'a\v{s}ek et al., 2008).

\subsection{Silicon mCP Star SX\,Arietis}

SX\,Ari~= 56\,Ari~= HD\,19832 is a fast rotating ($P\!=\!0\fd728$) Si mCP
star. Historically, it was the first mCP star, where the unsteady
period was revealed (Musie\l ok, 1998).

The behaviour of this star is complex: according to Adelman et
al.\ (2001) the secular rotational braking with a moderate rate of
2\,s/cen is superimposed over the cyclic variations of shapes and
amplitudes of light curves (\v{Z}i\v{z}\v{n}ovsk\'y et al., 2000;
Shore~\& Adelman, 1976) with a period of about five years, what
could be attributed to the precession of the rotational axis of a
magnetically distorted star.

On the basis of a precise Four College Photometric Telescope
$uvby$ photometry  several other mCP stars were revealed with
light curves also indicating precession (see the review paper by
Pyper~\& Adelman, 2004): e.\,g.\ 108\,Aqr (Adelman, 1997b),
20\,Eri (Adelman, 2000), V1093\,Ori (Pyper~\& Adelman, 2004),
MW\,Vul (Adelman~\& Young, 2005). As the changes of their periods
are marginal (if any), we do not include them among the stars with
unsteady rotation.

\subsection{Silicon CP Star EE\,Draconis}

The enigmatic EE\,Dra (HR\,7224~= HD\,177410) seemed to be a quite
ordinary fast--rotating Si CP star. Its rotational period, based
on the Winzer (1974), Hipparcos (ESA, 1998) and Adelman (1997a)
photometries is $P\!=\!1\fd1232$. This star, however, contrary to
the common magnetic CP stars, has not revealed a magnetic field,
which is very likely  due to its weakness (Krti\v{c}ka et al.,
2009; Shulyak et al., 2010).

\begin{figure}
\begin{minipage}[h]{0.5\linewidth}
\center{\vspace{4.5mm}\includegraphics[width=7.6cm,angle=0]{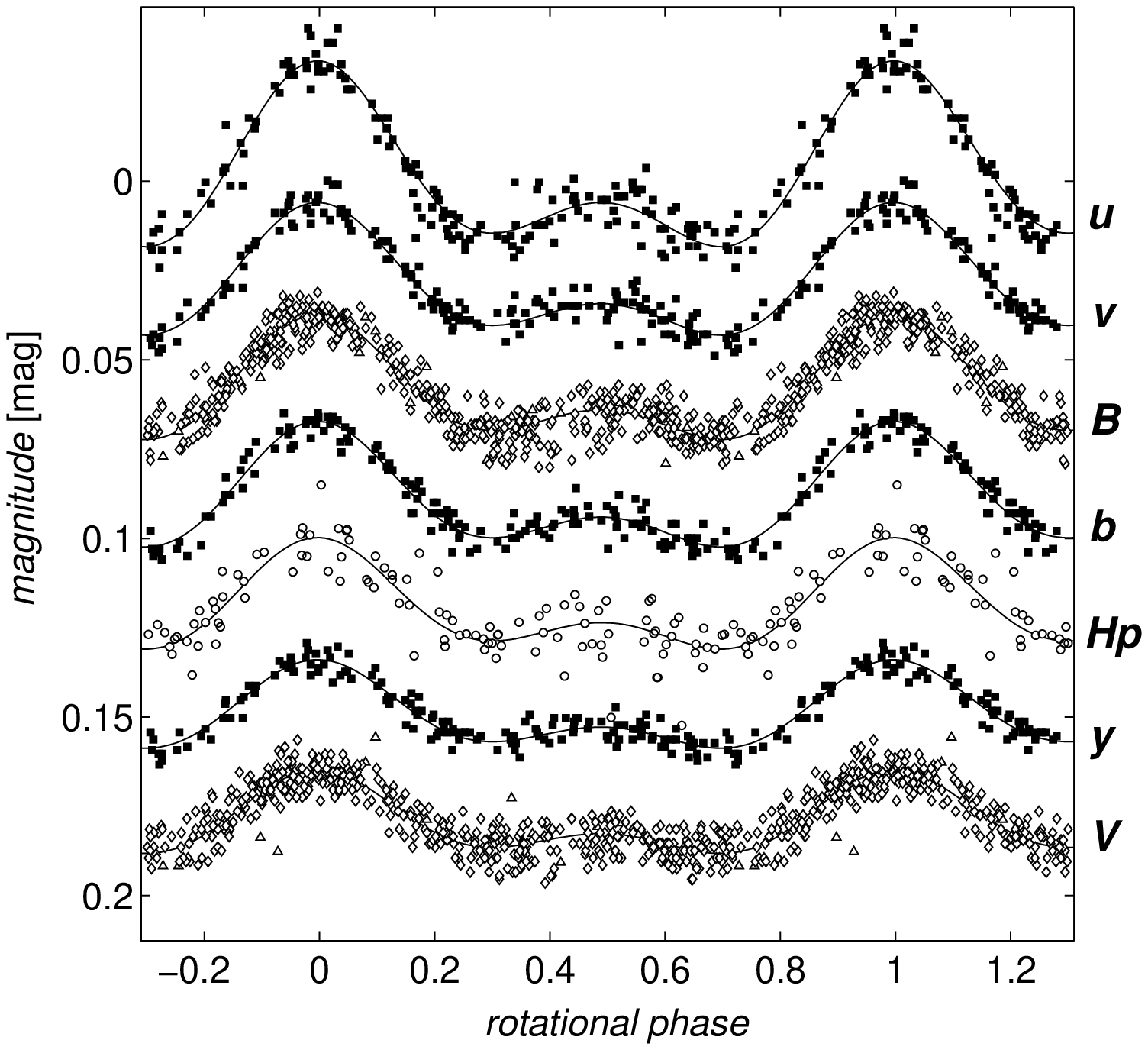}\\ (a)}
\end{minipage}
\begin{minipage}[h]{0.5\linewidth}
\center{\includegraphics[width=8.5cm,angle=0]{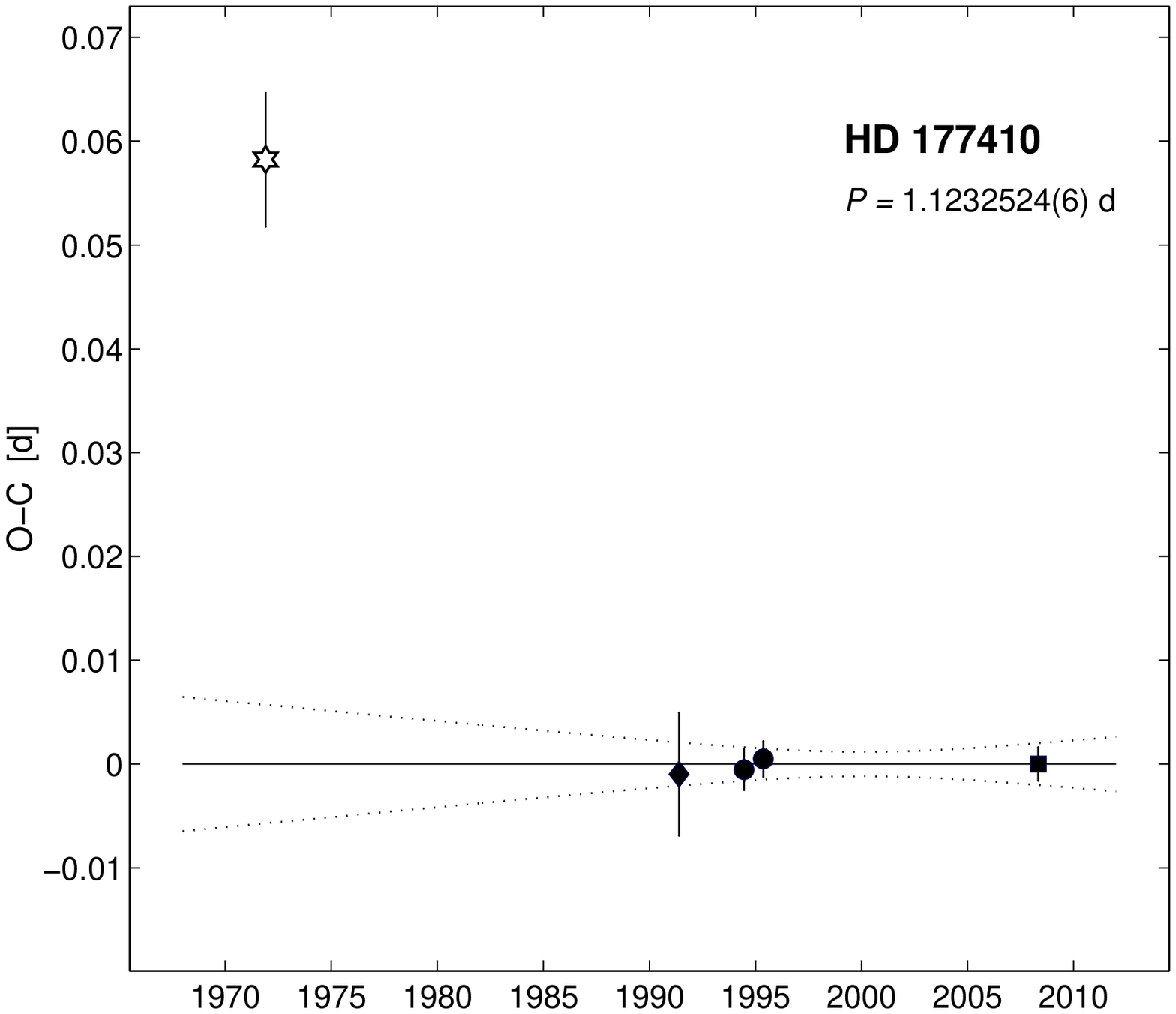}\\ (b)}
\end{minipage}
\caption{(a)~Light curves of EE\,Dra (HD\,177410) obtained in the
$\mathit{u,\ v,\ B,\ b,\ Hp,\ y}$ and $V$--bands. Observed changes
can be explained through uneven distribution of Si, Fe and other
chemical elements (for details see Krti\v{c}ka et al., 2009).
(b)~O--C diagram constructed from the times of light maxima,
derived from all the available photometry. The first O--C value
corresponds to the discussed observations of Winzer
(1974).}\label{OC_EE}
\end{figure}

The star exhibits a double--wave light curve (see
Fig.\,\ref{OC_EE}a) and strong variations of silicon lines.
Adelman (2004) reported an unprecedent rise of the amplitude of
the light variation from the typical 0.04 to 0.21\,mag and an
abrupt change of the period from the former $1\fd123$ to 101~days.
He attributed it to precession.

Later, Lehman et al.\ (2006, 2007) observed the star
spectroscopically and confirmed the period $1\fd1232$. Krti\v{c}ka
et al.\ (2009) then obtained new $BV$ photometry and refined the
period to $P\!=\!1\fd123524(6)$.

Not only Adelman's (2004) finding indicates an oddity of this
star, Winzer's (1974) data, which otherwise seems to be correct,
lie out the other ones, as can be seen on the O--C diagram,
Fig.\,\ref{OC_EE}b. Does this mean a quick lengthening of the
period between 1975 and 1990? Do the observations between 2003 and
2004 (Adelman, 2004) mean a reaction to the previous braking?

\section{Nature of Period Changes of mCP Stars}

The known chemically--peculiar stars with unsteady periods
represent a relatively diverse group; their O--C diagrams are
different, the common properties are rare, if any. It evokes a
situation when Edward Pigott (1753--1825) set up the first
catalogue of variable stars: it comprised only several objects,
but almost each of them represented other type of variability.
Similarly, the causes of the period instabilities of mCP stars may
be different.

\begin{table}
\caption{Summary of CP stars with unsteady periods, $P$~--- the
period, $\tau$~--- the spin--down time in Myr, $\Pi$~--- the
estimated duration of the cycle}
\begin{center}
\begin{tabular}{c|c|c|c|c|c|r}
  \hline\hline
  HD number & name & $P$ [d] & $\overline{\dot{P}}$ [s/cen] &
  $\dot{P}_{\rm{ex}}$ [s/cen]
  & $\tau$ [Myr] & $\Pi$ [yr] \\
  \hline\hline
  19832 & SX\,Ari & 0.728 & 2 & -- & 3 & $<\!250$ \\
  37776 & V901\,Ori & 1.539 & 53 & 66& 0.25 & $\sim\!90$ \\
  37479 & $\sigma$\,Ori\,E & 1.198 & 8 & -- & 1.3 & $<\!200$ \\
  124224 & CU\,Vir & 0.521 & 8 & 18 & 0.6 & $\sim\!60$ \\
  177410 & EE\,Dra & 1.123 & 1 & -- & 10 & $<\!500$ \\
  \hline
\end{tabular}
\end{center}
\end{table}

The spin--down time (SDT), $\tau\!=\!P/\overline{\dot{P}}$,
quantitatively represents the rate of the changing period of a star.
All the known cases of the period changes are positive (see Table~1)
implying braking of rotation, which implies that the process is
irreversible. Assuming that the SDT is constant, one can estimate
the maximum time--interval of the duration of the process (the
rotational period of the star cannot be shorter than the critical
one).

\subsection{Spin-Down or Cycle?}

Except for the extremely young $\sigma$\,Ori\,E the SDT values are
much shorter than the ages of the stars. Does it mean that the
rotational braking sometimes begins  long after the star arrives
at MS? Why? Why then do not we see a larger percentage of CP stars
with extremely long periods? Is it possible to brake the whole
star so drastically? Are the abrupt changes of the period of
CU\,Vir reported by Pyper et al.\ (1998); Pyper~\& Adelman (2004)
astrophysically permitted (the most dramatic case)? The last
question was brilliantly discussed by St\c{e}pie\'n (1998), who
clearly proved that one has to abandon the assumption of the
necessity of a rigid rotation and to admit that the outer layers,
controlled by magnetic field and denser inner parts can rotate
differently. This possibility was discussed and developed also in
Mikul\'a\v{s}ek et al.\ (2008, 2010).

The nature of CP stars leads us to the speculation about cyclic
variations of angular velocity  in the outer layers fixed by a
global magnetic field of several mCP stars. Let us assume the
simple sine course of such angular velocity variation with the
period $\Pi$. Then it is useful to introduce a parameter
$\Theta_{\rm{ex}}$ with the time dimension, where
$\Theta_{\rm{ex}}\!=\!\pi P\sqrt{2/|\dot{P}_{\rm{ex}}|}$. Here $P$
is the mean rotational period, $\dot{P}_{\rm{ex}}$ is the extremal
time derivative of the period (if known). Then the length of the
cycle $\Pi\!=\!\sqrt{\alpha}\,\Theta_{\rm{ex}}$, where $\alpha$ is
a dimensionless parameter expressing the amplitude of cyclic
changes in the O--C diagram in the units of  mean rotational
period.

Only two stars from the set of CP stars with unsteady periods,
discussed in the previous section have been observed for so long,
that we could estimate their maximum time derivatives of the
period $\dot{P}_{\rm{ex}}$: V901\,Ori, and CU\,Vir (see Tab.\,1).
After an inspection of their O--C diagrams we have accepted
$\alpha$ to be 0.5 as a first estimate. For V901\,Ori and CU\,Vir
with their $P\!=\!1\fd5387$, $\dot{P}_{\rm{ex}}\!=\!2.1\cdot
10^{-8}$ and $P\!=\!0\fd521$, $\dot{P}_{\rm{ex}}\!=\!5.7\cdot
10^{-9}$  we obtain the following estimates of duration of cycles
$\Pi$: 90 and 60 years, respectively.

In the case of other CP stars with changing periods we are forced to
manage with the estimate of the instant period derivative
$\overline{\dot{P}}$ (naturally, $\overline{|\dot{P}|}\!\leq\!
|\dot{P}_{\rm{ex}}|)$ we can introduce the similarly defined
parameter $\Theta$, $\Theta\!=\!\pi P\sqrt{2/|\dot{P}|}$, by means of
which we can estimate the maximum duration of the cycle of a
particular star: $\Pi\!\leq\!\sqrt{\alpha} \Theta$.

The cycle durations $\Pi$ for individual discussed CP stars are
given in Table~1. It appears that only for CU\,Vir and V901\,Ori
the expected cycles are short enough to observe them completely or
almost completely, the rotational periods of other stars are
changing too slow.

We can speculate that the thin outer envelope, dominated by the
global magnetic field, frozen in its plasma, performs with respect
to the inner part of the rotating star a torsional oscillation
along the rotational axis. Assuming that the oscillation period is
$\Pi$ with an amplitude of $A\!=\!2\pi\alpha R$, where $R$ is a
radius of the star. The maximum mutual equatorial velocity of the
oscillating envelope and the core is then
$v\!=\!4\pi^2R\alpha/\Pi$, the acceleration in turning points
$a\!=\!8\pi R\alpha/\Pi^2$. Numerically for $\Pi\!=\!75$ years,
$R\!=\!4\,R_{\odot}$ and $\alpha\!=\!0.5$ we get:
$A\!=\!13\,R_{\odot}$, $v\!=\!25$\,m/s and $a\!=\!7$\,m/s$^2$. The
torsion force should be connected with alternating protraction and
contraction of magnetic field lines. The process of the
oscillation excitation   is unclear. However, at the moment  the
speculations are only preliminary, and have to obtain a firmer
physical background.

\subsection{Where Are the Accelerating mCP Stars?}

If we assume the cyclic nature of the period variations, then we
should ask: ``Do any accelerating mCP stars exist?'' ``If yes, why
do we not see them?'' May be, at least one of the stars we
discussed is accelerating just now~--- the famous V901\,Ori~---
(see Fig.\,\ref{OC_901}b).

\begin{acknowledgements}
This work was supported by the grants: VEGA~2/0074/09, GA\v{C}R~205/08/0003,\\
MUNI/A/0968/2009, and the APVV project SK--CZ--0032--09.
\end{acknowledgements}

\end{document}